\documentclass[manuscript,showpacs]{revtex4}
\usepackage{graphicx}
\usepackage{amsmath}
\usepackage{amssymb}

\begin{document}

\title{Scheme for implementing atomic multiport devices}
\author{J.J. Cooper, D.W. Hallwood, and J. A. Dunningham}
\affiliation{School of Physics and Astronomy, University of Leeds, Leeds LS2 9JT, United Kingdom}

\begin{abstract}
Multiport generalizations of beam splitters are the key component in multipath interferometers, which are important in a range of quantum state engineering and precision measurement schemes.  Here we propose a straightforward method for implementing multiport devices for atoms trapped in optical ring lattices.  These  devices are interesting as atoms have certain properties (such as mass) that photons do not and the ring configuration makes them useful for applications such as precision gyroscopes. We discuss how they could be employed in useful measurement schemes and investigate how practical considerations limit the size of the devices that can be achieved by this method.
\end{abstract}

\maketitle

\section{Introduction}\label{introduction}

Multiport generalizations of beam splitters enable the manipulation of quantum states which may be important for quantum information processing on networks. They are also a key element in multipath interferometers and so have great potential in a range of measurement technologies.
Considerable progress has been made over the past decade towards implementing and understanding these devices.
Theoretical work has devised schemes to create multiport devices \cite{Reck1994a} and analyze some of their useful properties \cite{Vourdas2004a,Zhang2006a}. Experimental work using photons has demonstrated the operation of  beam splitters with three inputs and outputs (`tritters') as well as four inputs and outputs (`quarters') by clever arrangements of ordinary two-port beam splitters and phase plates \cite{Mattle1995a}. Similar devices have also been proposed \cite{Pryde2003a} and implemented \cite{Rasmussen94} in systems of bundled optical fibres. 

These devices become more powerful as the number of input and output ports is increased. However, for larger systems, the experimental configuration required rapidly becomes labyrinthine. In general, for a device with $S$ inputs and $S$ outputs, $S(S-1)/2$ beam splitters and $S(S+1)/2$ phase shifts would be required, i.e. a total of $S^{2}$ optical elements \cite{Reck1994a}. 
The complexity of implementation is perhaps even more problematic for atomic systems where beam splitting is a dynamic process typically involving the raising and lowering of potential barriers or the careful application of Bragg pulses. In this case, a multiport beam splitter would involve a complex sequence of operations and the experimenter would need to be able to address lattice sites individually -- an issue that has caused considerable problems when trying to use optical lattices for quantum information processing. 

In this paper, we demonstrate a scheme for implementing atomic multiport devices that goes someway towards overcoming these problems. In particular, it requires no additional equipment or steps than a standard two-port beam splitter. We begin by developing the theoretical scheme for an atomic tritter and follow with its extension to a device with a general number of ports. An interesting asymmetry is found in the behaviour of devices with an odd or even number of input ports. We then highlight how this scheme could be used for measurements.  Finally, we take into account various practical considerations and see that our scheme is likely to be limited to splitters with about five input (and output) ports.

\section{The scheme} \label{scheme}

The physical system we consider consists of an optical lattice of $S$ sites in a  `ring' configuration with atoms trapped at the potential minima. This can be described by the Bose-Hubbard model with Hamiltonian,
\begin{equation} \label{ham1}
H=\sum_{j=0}^{S-1}\epsilon_{j}a_{j}^{\dag}a_{j} -J\sum_{j=0}^{S-1}\left(a_{j}^{\dag}a_{j+1} + a_{j+1}^{\dag}a_{j} \right)+V\sum_{j=0}^{S-1}a_{j}^{\dag}{}^{2}a_{j}^{2},
\end{equation} 
where $a_{j}$ is the annihilation operator for an atom at site $j$ and the ring geometry means that $a_{S}=a_{0}$. The parameters $J$ and $V$ are the coupling and interaction strengths respectively, and $\epsilon_{j}$ accounts for the energy offset of site $j$. In general, we take the zero point energy to be the same for each site and so can ignore these energy offset terms. However, making one or more of the energy offsets non-zero for a fixed period of time can be a convenient way of imprinting phases on individual sites \cite{Denschlag03}. Initially we take the potential barriers between the sites to be sufficiently large that we can ignore tunnelling. This now is the starting configuration and each lattice site corresponds to an input port of our multiport device. 

The first step is to rapidly reduce the potential barriers between the sites in such a way that the sites still remain separate but are strongly coupled due to tunnelling. We want to do this rapidly with respect to the tunnelling time, but slowly with respect to the energies associated with excited states to ensure our system remains in the ground state. 
The adiabaticity criterion that ensures no excited states are populated is given, in general, by the phonon excitation spectrum $\{\omega_{k}\}$. This can be found using Bogoliubov theory and has the well-known form \cite{Oosten2001a, Rey2003a},
\begin{equation}
\omega_{k} = \sqrt{4J\sin^{2}\left(\frac{2\pi k}{S}\right)\left[ \frac{4NV}{S} + 4J\sin^{2}\left(\frac{2\pi k}{S}\right)\right]}, 
\label{equation}
\end{equation}
where $N$ is the total number of atoms, $S$ is the number of lattice sites, and the index $k$ runs over values $k = 0, 1, 2, ...., (S-1)$. Experiments have already successfully demonstrated this separation of timescales by ramping the optical intensity on a timescale of about $20$ms \cite{greiner2}. The limitations imposed by this condition are discussed further in Section~\ref{max}.

The coupling between wells now dominates over the interactions and the Hamiltonian (\ref{ham1}) can be written in the simple form,
\begin{equation}
H= -J\sum_{j=0}^{S-1} \left(a_{j}^{\dag}a_{j+1} + a_{j+1}^{\dag}a_{j} \right). \label{hamaux}
\end{equation}

Of course this makes the approximation that the interaction energy will be negligible compared with the coupling energy. In practice, the interactions will not be strictly zero. However, they can be made very small with respect to $J$ by, for example, making use of Feshbach resonances to tune the scattering lengths \cite{feshbach}. For now, it is helpful to consider the case where we can ignore $V$. We consider the effects of interactions in Section \ref{max}.

\subsection{Example: The Tritter} \label{sec:tritter}

\begin{figure}[h] 
\includegraphics[scale=1]{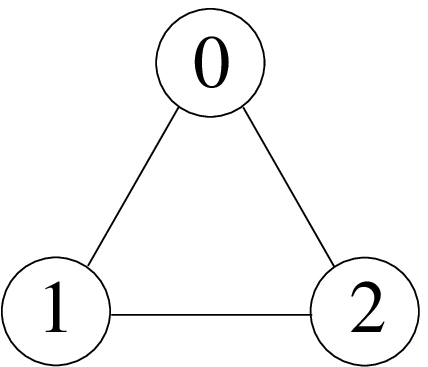}
\caption{Ring configuration of the sites in the optical lattice for the tritter. The lines denote coupling between sites due to tunnelling.}
\label{ringfig}
\end{figure}

We begin by considering  the case of three lattice sites in a ring (see Figure \ref{ringfig}).
A similar configuration has been achieved experimentally by trapping atomic Bose-Einstein condensates (BECs) in the optical potential created by the diffraction of a laser beam by a liquid crystal spatial light modulator \cite{Boyer2006a}. This modulator allows arbitrary three-dimensional trapping potentials to be achieved, which have the added advantage of being able to be varied smoothly with time. 
Another promising possibility for creating the ring potential, is to interfere a Laguerre-Gaussian (LG) laser beam with a plane wave co-propagating along the $z$-direction 
\cite{Amico2005a}. By retro-reflecting this combined beam, a standing wave can be formed that consists of a stacked array of disk shaped traps along the $z$-direction. By 
controlling the tunnelling between the disks and making it much smaller than the 
corresponding tunnelling within each ring, one can implement an array of effective 
1-D ring lattices. In both these cases, the rate of tunnelling between the sites in a ring can be 
controlled simply by adjusting the intensity of the trapping laser light.

The Hamiltonian (\ref{hamaux}) describing this system is,
\begin{equation} \label{ham2}
H= -J\left( a_{0}^{\dag}a_{1} + a_{1}^{\dag}a_{2} + a_{2}^{\dag}a_{0}\right) + \mbox{h.c.}
\end{equation}
This can be diagonalized in the basis:
\begin{equation}
\left(\begin{array}{c}
\alpha_{0}\\ \alpha_{1}\\ \alpha_{2}
\end{array}\right) =
\frac{1}{\sqrt{3}} \left( \begin{array}{ccc}
1 & 1 & 1 \\
1 & e^{i2\pi/3} & e^{-i2\pi/3} \\ 
1 & e^{-i2\pi/3} & e^{i2\pi/3} \end{array}\right)
\left(\begin{array}{c}
a_{0} \\  a_{1} \\  a_{2}
\end{array}\right) \equiv U \left(\begin{array}{c}
a_{0} \\  a_{1} \\  a_{2}
\end{array}\right)
\end{equation}
to give,
\begin{equation}
H = -J\left[ 2\alpha_{0} ^{\dag}\alpha_{0} - \alpha_{1} ^{\dag}\alpha_{1}  - \alpha_{2} ^{\dag}\alpha_{2} \right].
\end{equation}
If the system is allowed to evolve for time, $t$, the $\alpha_{0}$ mode acquires a phase of $-2Jt$, while the $\alpha_{1}$ and $\alpha_{2}$ modes each acquire a phase of $Jt$. If the barriers are then raised on a similar timescale to their lowering (i.e. quickly with respect to the tunnelling time, but slowly with respect to the energies associated with excited states), the atoms are `frozen' in the lattice sites $a_{0}$, $a_{1}$, and $a_{2}$. The overall output operators $\{ A_{0}, A_{1},A_{2}\}$ are given by:
\begin{equation}
\left(\begin{array}{c}
A_{0} \\  A_{1} \\  A_{2}
\end{array}\right) = U^{-1}\left( \begin{array}{ccc}
e^{i2Jt }& 0 & 0 \\
0 & e^{-iJt} & 0 \\ 
0 & 0 & e^{-iJt} \end{array}\right) U \left(\begin{array}{c}
a_{0} \\  a_{1} \\  a_{2}
\end{array}\right) 
=  \frac{1}{3}
\left( \begin{array}{ccc}
\Omega_{1}& \Omega_{2} & \Omega_{2} \\
\Omega_{2}& \Omega_{1} & \Omega_{2} \\
\Omega_{2}& \Omega_{2} & \Omega_{1} \\
\end{array}\right)  \left(\begin{array}{c}
a_{0} \\  a_{1} \\  a_{2}
\end{array}\right)
\equiv R_{3}\left(\begin{array}{c}
a_{0} \\  a_{1} \\  a_{2}
\end{array}\right),          \label{trittertrans}
\end{equation}
where $\Omega_{1} = e^{i2Jt}+2e^{-iJt}$, $\Omega_{2} = e^{i2Jt} - e^{-iJt}$, and for later reference, we have defined the overall operation as $R_{3}$. 
We see that the output modes are identical to the input modes when $t=0$, as we would expect. 
For a balanced tritter, we need the output modes to be an equal superposition of the input modes, i.e. $|\Omega_{1}| = |\Omega_{2}|$. This occurs when $t=2\pi/9J$, for which value we have (ignoring any irrelevant overall phase),
\begin{equation}
R_{3} = \frac{1}{\sqrt{3}}\left( \begin{array}{ccc}
1& e^{i2\pi/3} & e^{i2\pi/3}  \\
e^{i2\pi/3}& 1 & e^{i2\pi/3} \\
e^{i2\pi/3} & e^{i2\pi/3} & 1 \\
\end{array}\right). \label{R3def}
\end{equation}

To summarise, the steps in implementing a balanced tritter are:
\begin{enumerate}
\item{Rapidly reduce the potential barriers separating the lattice sites.}
\item{Allow the system to evolve for time $t=2\pi/9J$.}
\item{Rapidly raise the potential barriers.}
\end{enumerate}

Each output is an equally-weighted superposition of all the inputs. One may notice, however, that the phases between terms in these outputs are different from those commonly quoted in the literature (e.g. in \cite{Mattle1995a, Vourdas2004a}). 
For many purposes, this does not matter. For example, we can take the output modes of a 50:50 beam splitter to be either $\{(a_{0}+a_{1})/\sqrt{2}, (a_{0}-a_{1})\sqrt{2}\}$ or $\{(a_{0}+ia_{1})/\sqrt{2}, (ia_{0}+a_{1})\sqrt{2}\}$
without fundamentally changing the results.
If, however, we do require a particular form of the phases for the output modes, this can always be achieved by imprinting phases on individual lattice sites before and after the lowering and raising of the barrier. In practice this could be achieved by applying energy offsets, $\epsilon_{j}$, to the lattice sites for some fixed time, as discussed above. 

This procedure not only achieves the goal of a multiport beam splitter for atoms, but is much simpler than schemes that combine phase shifts with a complicated network of beam splitters. Our scheme only requires a lowering and raising of the lattice potential and, importantly, requires no more operational effort than an ordinary beam splitter for atoms  \cite{Dunningham2004a}. This bodes well for the possibility of scaling the scheme up to larger systems.

\section{Larger devices}\label{larger}
We now extend this scheme to the general case of a lattice ring with an arbitrary number of sites, $S$ (in Section \ref{max} we will determine up to what values of $S$ this scheme is valid experimentally).  Following the same procedure as for the tritter we lower the potential barriers so that coupling between the wells dominates over the interactions and the Hamiltonian describing the system is given by (\ref{hamaux}).  This can be written in the diagonalised basis as,
\begin{equation}
H = -2J \sum_{k=0}^{S-1} \cos \left(\frac{2 \pi k}{S}\right) \alpha_{k}^{\dagger} \alpha_{k}
\label{hamS}
\end{equation}
where,
\begin{equation}
\alpha_k = \frac{1}{\sqrt{S}}\sum_{j=0}^{S-1} e^{i2\pi jk/S}a_{j}.
\label{fourierS}
\end{equation}
The system is now allowed to evolve for time $t$ during which the $S$ modes each acquire a phase of $-2 \cos \left(\frac{2 \pi k}{S}\right)Jt$.   The potential barriers are then quickly raised and the output is given by,
\begin{align}
\begin{pmatrix}
A_{0} \\
A_{1} \\
\vdots \\
A_{S-1}
\end{pmatrix} &= U^{-1}
\begin{pmatrix}
e^{i2Jt} & 0 & 0 \ldots & 0 \\
0 & e^{i2 \cos \left(\frac{2 \pi }{S}\right)Jt} & 0 \ldots & 0 \\
\vdots & \vdots & \ddots & \vdots\\
0 & 0 & \ldots & e^{i2 \cos \left(\frac{2 \pi (S-1)}{S}\right)Jt}
\end{pmatrix}
U
\begin{pmatrix}
a_{0} \\
a_{1} \\
\vdots \\
a_{S-1}
\end{pmatrix}  \nonumber \\
\nonumber\\
&=
\frac{1}{S}
\begin{pmatrix}
\Omega_{1} & \Omega_{2} & \ldots & \Omega_{S} \\
\Omega_{S} & \Omega_{1} & \ldots &  \Omega_{S-1} \\
\vdots & \vdots & \ddots & \vdots \\
\Omega_{2} & \Omega_{3} & \ldots & \Omega_{1}
\end{pmatrix}
\begin{pmatrix}
a_{0} \\
a_{1} \\
\vdots \\
a_{S-1}
\end{pmatrix}
\equiv R_{S}
\begin{pmatrix}
a_{0} \\
a_{1} \\
\vdots\\
a_{S-1}
\end{pmatrix}
\label{generals}
\end{align}
where the unitary matrix is given by $U_{kj}= \frac{1}{\sqrt{S}} e^{i2\pi jk/S}$. 

As before, the condition for a balanced multiport splitter is that each output is an equally weighted superposition of all the inputs, i.e. $|\Omega_1| = |\Omega_2| = \ldots = |\Omega_S|$.  We now wish to see if our procedure can achieve this for a general $S$ ignoring, for now, factors that may limit its experimental feasibility.

\subsection{Producing a balanced splitter}
\begin{figure}[h]
\centering
\includegraphics[scale=0.5]{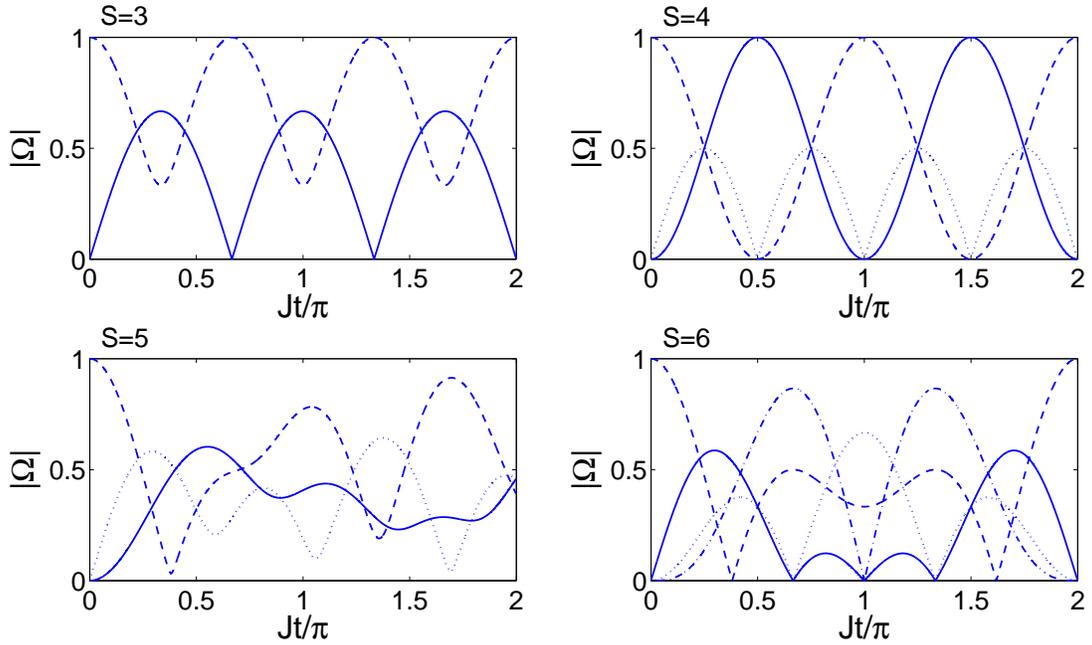}
\caption{These plots show how the $\Omega$ vary with time for multiport devices with $S=3,4,5$ and $6$ from top left to bottom right respectively.  We see that for $S=3$ the $\Omega$ intersect at $t=2\pi /9J$ and for $S=4$ they intersect at $t=\pi /4J$.  For $S=5$ and $6$ there is no intersection of all the $\Omega$ in the time range, $t=0 \to 2\pi /J$, shown here.}
\label{omega3456}
\end{figure}

We can determine if a balanced splitter can be achieved for a device with $S$ sites by plotting the elements of $R_{S}$ as a function of $Jt$ and checking whether there are times for which $|\Omega_1| = |\Omega_2| = \ldots = |\Omega_S|$.  Figure \ref{omega3456} shows these plots for $S=3,4,5$ and $6$ respectively in the time range $t=0 \to 2\pi /J$.  We see that all the $\Omega$ intersect at $t=2\pi/9J$ for $S=3$ and at $t=\pi/4J$ for $S=4$, but there is no exact crossing for $S=5$ and $6$ in this range (or indeed for larger values of $S$).  However, looking in the time range $t=0 \to 5000/J$ we find for systems with odd values of $S$, up to $S=9$, there is an intersection of all the $\Omega$ to within an error of $1\%$. 
It is convenient to define a measure, $\chi$, of how good the intersection is, as
\begin{equation}
\chi=\sum_{i=1}^{S}\left(\left|\Omega_{i}\right| - \frac{1}{\sqrt{S}} \right)^{2} . \label{def:chi}
\end{equation}
This is simply the sum of the squares of the deviation of each value of $|\Omega|$ from its ideal value (i.e $1/\sqrt{S}$) -- as discussed above, we are not concerned by the phase of the $\Omega$ values. If, on average, each value of $|\Omega|$ differs from $1/\sqrt{S}$ by $1\%$, $\chi$ has a value of $10^{-4}$. We will use this value (i.e. $\chi \lesssim 10^{-4}$) as a useful criterion for when a balanced  splitter has been achieved.

\begin{figure}[h]
\centering
\includegraphics[scale=0.4]{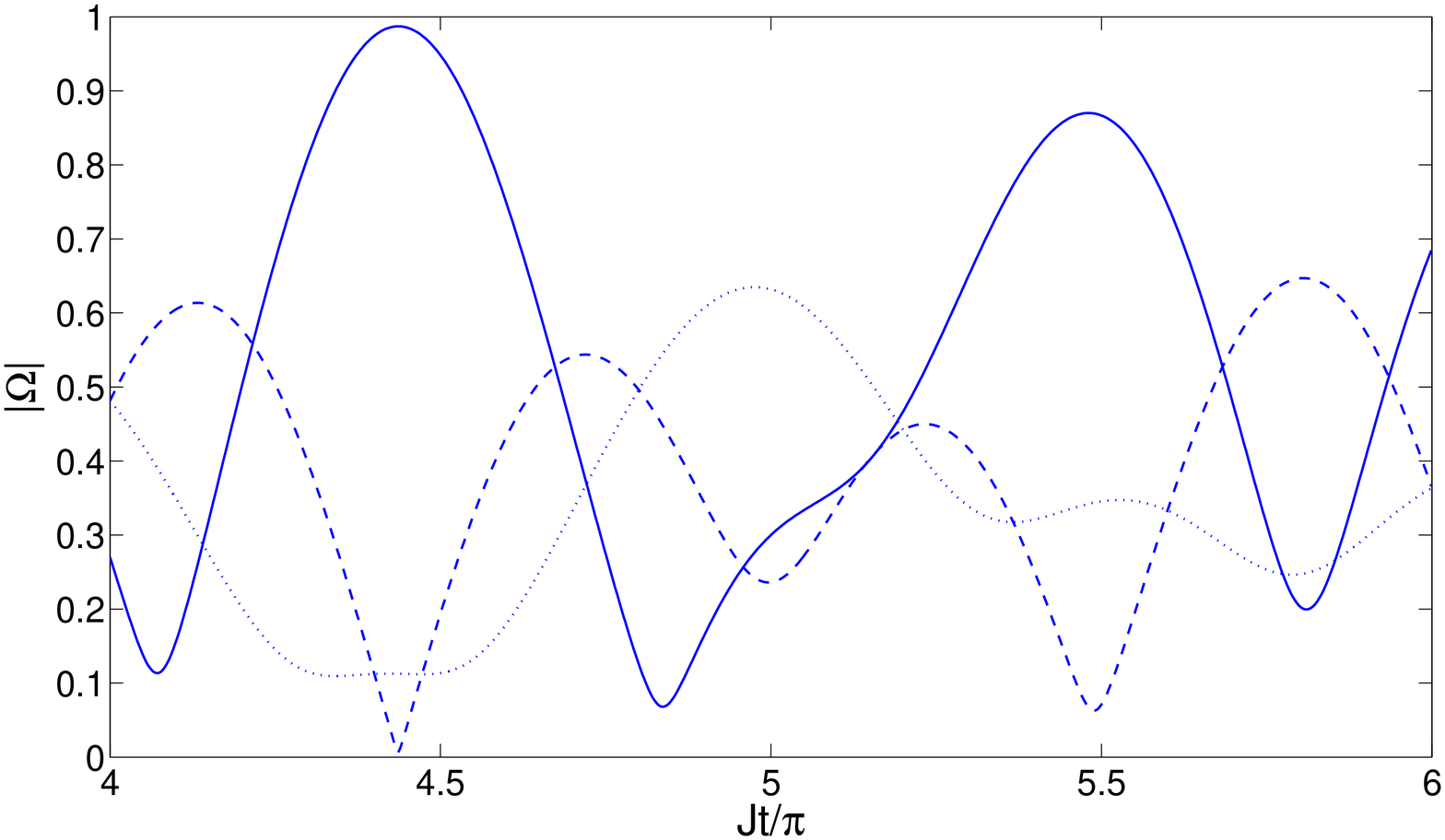}
\caption{Using $\chi \lesssim 10^{-4}$ as our defining criterion for when a balanced splitter is achieved we find a five site splitter is produced at $t \approx 5.2\pi/J$.  The near intersection of all the $\Omega$ at this time can be seen here.}
\label{omega5_f2}
\end{figure}

Using this criterion, we find balanced splitters for $S=5, 7$ and $9$ at evolution times $t\approx 5.2 \pi/J$, $88\pi/J$ and $177\pi/J$ respectively.  This intersection for $S=5$ is shown in Figure \ref{omega5_f2}.  The time required becomes longer because there are more values of $\Omega$ to match.  Interestingly,  no intersections for which $\chi \lesssim 10^{-4}$ are achieved for even values of $S$ in this time range. We will discuss possible reasons for this in the next section.  We should note that while we do not see intersections that satisfy $\chi \lesssim 10^{-4}$ for even values of $S$ in the time range used, this does not rule out the possibility at a later time. One exception is $S=6$ because all the values of $\Omega$ in this case repeat with a period of $2\pi$ and, since there is no intersection during this period, there will never be one.  The same may be true for some other, larger $S$.  However, so long as $\chi$ for a given $S$ is not periodic in time, if we wait long enough the balanced splitting criterion should be able to be achieved. Of course, this may not always be experimentally expedient and could be longer than the coherence time of the system.

To summarise, devices with even and odd numbers of sites are potentially both capable of producing balanced multiport devices.  However, the even cases usually require a much longer evolution time than the odd ones.  We now investigate the reasons for this difference in behaviour.

\subsection{Evolution while the barriers are lowered}

For simplicity we will begin by considering the evolution of a state containing only a single particle. However, our results also apply to more general states. An initial single particle state can be written as,
\begin{equation}
|\Phi\rangle = \sum_{j=0}^{S-1}A_{j}a_{j}^{\dagger}|0\rangle = \sum_{k=0}^{S-1}B_{k}\alpha_{k}^{\dagger}|0\rangle ,
\label{AjBk}
\end{equation}
where $A_{j}$ and $B_{k}$ are the amplitudes in the two bases $a_j$ and $\alpha_k$ respectively, and are related by a Fourier transform.  Using (\ref{hamS}), (\ref{fourierS}) and (\ref{AjBk}) we can write the evolution of the state while the barriers are lowered as,
\begin{equation}
|\Phi (t)\rangle = \frac{1}{\sqrt{S}} \sum_{k,j=0}^{S-1}B_{k}\exp\left[i2Jt \cos\left(\frac{2\pi k}{S}\right) + \frac{i2\pi kj}{S}\right]a_{j}^{\dagger}|0\rangle .
\end{equation}
Taking the case where the single atom is initially in site $0$, i.e. $A_{0}=1$ and $A_{n}=0$ for $n=1,2, \ldots , S-1$, for an even value of $S$ we can write,
\begin{equation}
|\Phi(t)\rangle=\sum_{j=0}^{S-1}e^{i\pi j/2}\sum_{k=1}^{S/4-1}\Lambda(j,k,S,t)\, a_{j}^{\dagger}|0\rangle
\label{phiteven}
\end{equation}
where $\Lambda(j,k,S,t)$ is some real-valued function that depends on $j$, $k$, $S$ and $t$.  This can be written more simply as,
\begin{equation}
|\Phi(t)\rangle= \sum_{j=0}^{S-1} \tilde{\Lambda}(j,S,t)\, e^{i\pi j/2}\,a_{j}^{\dagger}|0\rangle,
\label{phiteven2}
\end{equation}
where $\tilde{\Lambda}(j,S,t)$ is also a real-valued function. 
Importantly from (\ref{phiteven2}) we see that adjacent sites always have a phase difference of $\pm \pi/2$.  This means that the rate at which atoms can flow between sites has certain fixed values since the flow velocity is given by, $v=(\hbar /m) \nabla \phi$ where $\nabla \phi$ is the phase gradient or phase difference.

By contrast, for an odd number of sites the wave function after evolution for time $t$ can be written as,
\begin{eqnarray}
|\Phi(t)\rangle &=& \sum_{j=0}^{S-1}\left(e^{i2Jt}+e^{i\pi j/2}\sum_{k=1}^{S/4-1}\Lambda(j,k,S,t)\right)\,a_{j}^{\dagger}|0\rangle \nonumber \\
&=& \sum_{j=0}^{S-1}\left(e^{i2Jt}+e^{i\pi j/2}\,\tilde{\Lambda}(j,S,t)\right)\,a_{j}^{\dagger}|0\rangle .
\label{oddphase}
\end{eqnarray}
Here the phase at each lattice site varies continuously with time and so the phase differences between adjacent sites, and hence the velocity of flow, are not constrained in the same way as the even case. The additional constraint imposed by the symmetry of the even case on the allowed flow rates is one reason why it is more difficult for devices with an even number of sites to achieve a state where their outputs are in an equal superposition of their input modes \footnote{This argument suggests that it should be difficult to achieve a balanced splitter for $S=4$, but we have seen that this is not the case.  This is because the constraint that adjacent sites have a phase difference of $\pm \pi/2$ happens to be precisely the  phase difference required for the $S=4$ case to work}. We have taken the simple example of a single particle state, however the same argument holds for general input states: the only change is the form of the real-valued functions.

Another difference between the cases of even and odd numbers of sites can be found by considering the direction of the flow of atoms around the ring.  In the case of even $S$ the amplitude of each site is given by $e^{i\pi j/2}\tilde{\Lambda}(j,S,t)$, and so the phase is always $\pm 1$ or $\pm i$. The function $\tilde{\Lambda}(j,S,t)$ is smoothly varying and so changes sign only when it passes through zero. Consequently the phase difference between two sites, and hence the direction of flow, changes only when the amplitude of one of the sites is zero. The same is not true for the devices with odd $S$ as their phases change continuously with time.  This allows the direction of flow between the sites to change smoothly with time.  This difference in the evolution of systems with an even number of sites and those with an odd number of sites again suggests it will be  more difficult for those with even $S$ to achieve the required balanced output.

The final difference is seen when we look at the number of distinct elements required to compose the operators, $R_{S}$. Instead of the number of different $\Omega$s equalling the number of sites, $S$, there are $(S+1)/2$ different $\Omega$s required for odd $S$ and $(S+2)/2$ different $\Omega$s for even $S$.  This is due to the symmetry of the system and it means for even $S$ there are more values of $\Omega$ to match to make a balanced splitter.  Again, this difference supports our observation that it is more difficult for systems with an even number of sites to produce a balanced splitter.

\section{Inverse transforms and unbalanced splitters}\label{inverse}

Now that we have understood the basic operation of multiport splitters, we would like to consider how they can be combined to create useful quantum devices. One simple, yet important, such device is an interferometer. 
In general, an interferometer consists of a beam splitter, a phase shift, and then an inverse beam splitter.  In standard two-path interferometry, the inverse beam splitter is replaced with a normal beam splitter. The reason this works is that a 50:50 beam splitter can be thought of as a $\sqrt{\rm NOT}$ operation. 
We can see this because a Mach-Zehnder interferometer with no phase difference between the two paths (i.e. two 50:50 beam splitters in succession) gives an output state that is the same as the input but with the ports swapped, i.e. a ${\rm NOT}$ operation.
So two beam splitters in succession will return us to the original state so long as we make a trivial swap of the labels of the output ports. For multipath interferometers, however, the same is not true: we cannot simply replace the inverse splitter with an ordinary splitter and relabel the modes.   So, to be able to implement a multipath interferometer, we need to be able to implement inverse multiport devices.

It turns out that {\em three} successive applications of a tritter, $R_{3}^{3}$ is equivalent to the identity and leaves the original input state unchanged. This can easily be verified using Equation (\ref{R3def}). This means that the inverse operation of a tritter, $R_{3}$, is simply $R_{3}^{2}$ or, equivalently, a tritter operation where the state is allowed to evolve for twice as long with the barriers lowered (i.e. $t=4\pi/9J$ rather than $t=2\pi/9J$). The inverse operation can, therefore, be implemented just as easily as the original transform.

More generally, the inverse operation of a device with $S$ sites is simply $R_{S}^{(S-1)}$. We have numerically checked this for $S=4, 5, 7$, and $9$ \footnote{$S=6$ and $S=8$ were not considered due to the difficulty in obtaining a splitter in these cases, as discussed above} by calculating how close $R_S R_{S}^{(S-1)} = R_{S}^{S}$ is to the identity operator. A useful measure of this is simply to sum up the modulus squares of the leading diagonal of $R_{S}^{S}$ and divide by $S$ since this effectively gives the fidelity of the output state relative to the input (averaged over all input ports). The values we found for this measure for $S=4,5,7$ and $9$ were $1, 0.96, 0.91$ and $0.75$ respectively.  We see that the inverse splitter is degraded as $S$ is increased. This is not surprising because, if our timing does not give an exact balanced splitter, when we multiply this time by $(S-1)$ to give the time of the inverse splitter the imperfection is multiplied. 

So far, we have only discussed splitters that have balanced outputs and are therefore the multiport generalizations of 50:50 beam splitters. For many applications we may not want the outputs to be balanced. For example, we may want a device that coherently skims off only a small fraction of an input state and redistributes it between the output modes.  Our scheme holds great potential for producing such devices simply by changing the value of the evolution time. Since the coherent amplitude in each output mode depends on $t$, this allows us to obtain multiport splitters with different `reflectivities'.
All this can be achieved without changing the experimental set-up -- only the timings of the steps.  It is important to note, however, that not all unbalanced outputs can be achieved by this method. For example, in Figure~\ref{omega3456} we see that not all the values of $\Omega$ are independent.

\begin{figure}[h] 
\includegraphics[width=10.0cm]{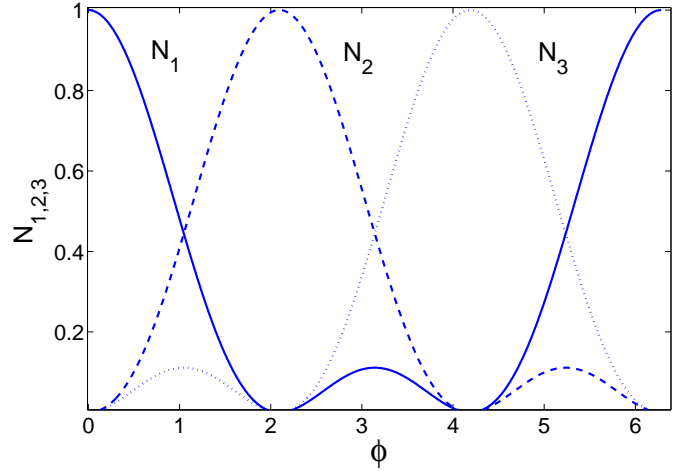}
\caption{The fraction of the total number of atoms, $N_{1}$, $N_{2}$ and $N_{3}$, in each of the three output ports of a three-path interferometer given by Eq.~(\ref{threepathout}). The applied phase, $\phi$, inside the interferometer is the same across each pair of adjacent sites.} 
\label{interferometer}
\end{figure}

We now consider how these multiport splitters can be combined into useful devices. A simple, but illuminating, example is that of a three-path interferometer. Suppose we have three lattice sites in a ring configuration and start with $N$ atoms in one site, $a_{0}$. The interferometer is implemented by a tritter operation, $R_{3}$, a phase shift on the lattice sites, and then an inverse tritter, $R_{3}^{-1}$. If we take the case that there is a linearly varying phase shift of $\phi$ between adjacent lattice sites the overall transformation of the interferometer is,
\begin{equation}
\left(\begin{array}{c}
A_{0} \\  A_{1} \\  A_{2}
\end{array}\right) = R_{3}^{-1}\left( \begin{array}{ccc}
1 & 0 & 0  \\
0 & e^{i\phi} & 0  \\ 
0 & 0 & e^{i2\phi}  
 \end{array}\right) R_{3} \left(\begin{array}{c}
1 \\  0 \\  0  
\end{array}\right).
\end{equation}
The mean number of atoms at each final lattice site (normalized by the total number) is then given by:
\begin{equation}
\left(\begin{array}{c}
N_{0} \\  N_{1} \\  N_{2}
\end{array}\right) = \frac{1}{9}\left( \begin{array}{c}
3+4\cos(\phi) + 2\cos(2\phi)   \\
3+4\cos(\phi-2\pi/3)+ 2\cos(2\phi+2\pi/3)  \\ 
3+4\cos(\phi+2\pi/3)+ 2\cos(2\phi-2\pi/3)
 \end{array}\right). \label{threepathout}
\end{equation}
These mean numbers of atoms are plotted as a function of $\phi$ in Figure \ref{interferometer}. As we might expect from such an interferometer, a measurement of the final population in each of the three lattice sites allows us to uniquely determine the value of $\phi$ (modulo $2\pi$).
Importantly, we note that this three-path interferometer requires no more effort than a two-path scheme for atoms (and indeed a scheme with many paths would also require no additional equipment or steps). This is in contrast with existing schemes \cite{Reck1994a} that combine beam splitters in complicated arrays and illustrates the power of this scheme.

\section{Practical limitations to the number of ports}\label{max}

So far in discussing the general scheme for implementing an atomic multiport device we have neglected a number of important physical processes. This has allowed us to treat an idealised case and has been useful for understanding its key features. However, in reality, these processes must be accounted for and will limit the applicability of the scheme. In this Section we highlight the key practical limitations to this scheme and assess their impact.

\subsection{Condensate lifetime}
\begin{figure}[h]
\centering
\includegraphics[scale=0.4]{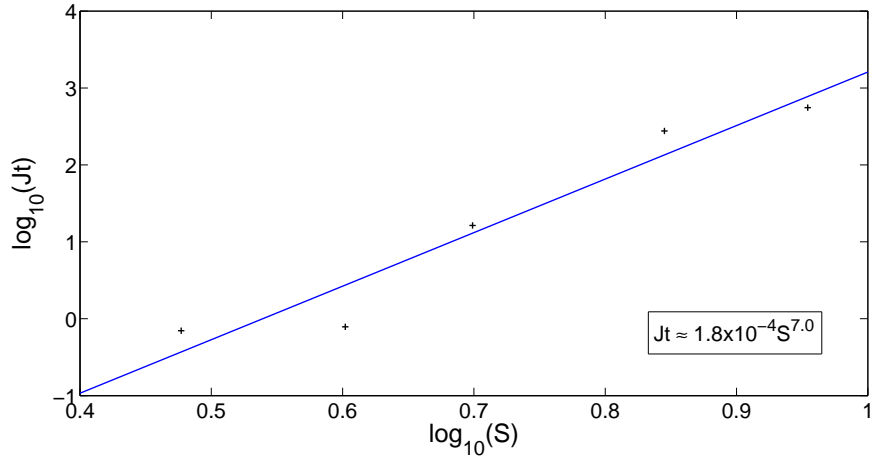}
\caption{The $Jt$ required to produce a balanced splitter for different $S$.  We find these $Jt$ scale with $S$ as $1.8$x$10^{-4}S^{7.0}$.}
\label{Sscaling}
\end{figure}

One potential limiting factor is how long the system must be allowed to evolve to produce the balanced splitters since this may be longer than the lifetime of the condensate.  In Section \ref{larger} we showed that the evolution times increased with $S$.  Figure \ref{Sscaling} shows the value of $Jt$ required to produce balanced splitters for $S=3, 4, 5, 7$ and $9$ and we see that in this range $Jt$ scales as approximately the seventh power of $S$.  This is quite a prohibitive scaling and in order to determine what values of $S$ are feasible, we need to compare the evolution time with the lifetime of the condensate.

Condensate lifetimes of about 10s have been measured \cite{Esteve08} and oscillation periods of 40ms have been observed between wells in bosonic Josephson junctions \cite{Albiez05}, giving $J \sim 80$s${}^{-1}$.  Together these give an estimate of a typical maximum value of $Jt\approx 800$. For $S=3, 4, 5, 7$ and $9$ the largest value of $Jt$ is $177\pi$ which is comfortably below 800.  Therefore, the times required to produce balanced splitters with up to nine sites should all be experimentally achievable.  Using the scaling of $Jt$ with $S$ above, for $S=11$, $Jt>800$ suggesting $S=9$ is the maximum number of sites the evolution times allow.

\subsection{Intensity fluctuations} \label{intensity}
Since the time for which the barriers must be lowered depends on the tunnelling rate, any uncertainty in $J$ will introduce imperfections into the scheme. This may concern some readers since it is well known that $J$ depends exponentially on the intensity of the trapping light which is subject to fluctuations. 
However, just because $J$ depends exponentially on the intensity does not mean that small fluctuations in the intensity will result in large fluctuations in the tunnelling rate. Indeed the exponential function varies only linearly for exponents close to zero, and this is the regime in which this scheme operates. 
We can check this using an approximation for $J$ found in \cite{Scheel06},
\begin{equation}
\hbar J \approx \frac{E_R}{2}\exp \left(\frac{-\pi^2}{4} \sqrt{\frac{V_0}{E_R}}  \right) \left[\sqrt{\frac{V_0}{E_R}} + \left(\sqrt{\frac{V_0}{E_R}} \right)^3 \right]
\label{J}
\end{equation}
where $E_{R}$ is the atomic recoil energy, $V_{0}$ is the potential depth and $\hbar=1$ in \cite{Scheel06}.  To ensure we are in the strong coupling regime we take $V_{0}=2E_{R}$, a configuration that has been achieved experimentally \cite{Jona03}.  For a small fractional fluctutation, $\delta$ in the exponent due, for example, to intensity fluctuations in the trapping laser, this can be rewritten as,
\begin{eqnarray*}
\tilde{J} \approx \exp \left(\frac{-\pi ^ 2}{2\sqrt{2}}(1+\delta)\right)  = J \exp \left(\frac{-\pi ^{2}}{2\sqrt{2}} \delta \right).
\end{eqnarray*}
Intensity fluctuations can be stabilised to around $0.1\%$ \cite{Kwon03}, i.e. $\delta = 0.001$, which gives $\tilde{J} \approx 0.9965 J$. So the uncertainty in the tunnelling rate due to intensity fluctuations is modest and unlikely to be the major limitation to this scheme.

\subsection{Interactions}
Up until now we have neglected the interactions between atoms by setting $V=0$.  We shall now consider how non-zero interactions limit our scheme.  Taking them into account the Hamiltonian describing the system is,
\begin{equation}
H=-J\sum_{j=0}^{S-1}\left(a_j^{\dagger}a_{j+1} + a_{j+1}^{\dag}a_{j} \right)+ V\sum_{j=0}^{S-1}a_j^{\dagger 2}a_j^{2}.
\label{perturbation}
\end{equation}
First, we shall consider the effect of these interactions on our tritter for different numbers of atoms.  We will then determine how this effect scales with $S$.

\begin{figure}[h]
\centering
\includegraphics[scale=0.4]{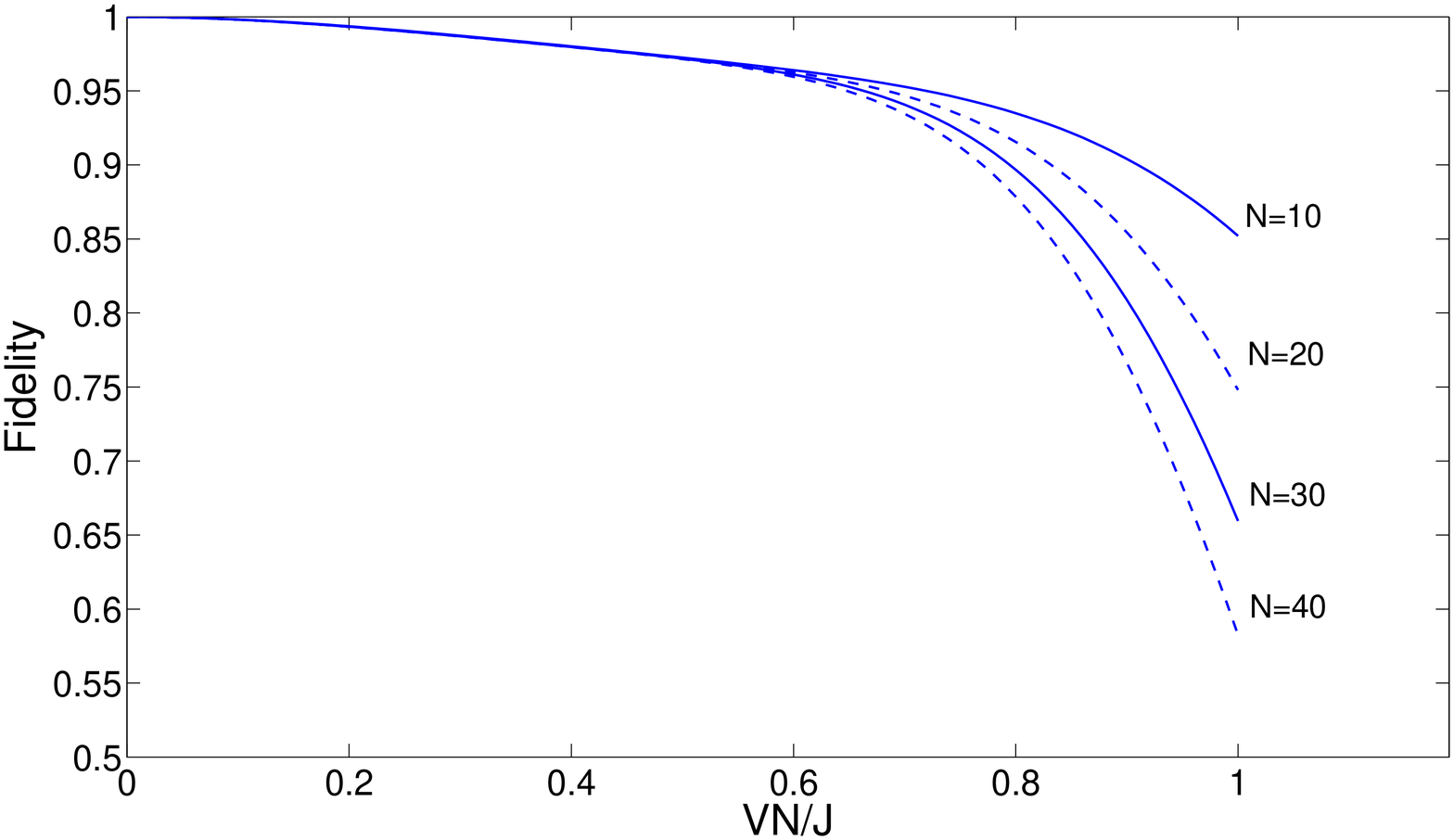}
\caption{The fidelity of the output states of a tritter with interactions (calculated from the overlap with the output when there are no interactions) plotted over the range $VN/J = 0 \to 1$ for different numbers of atoms.}
\label{fidn}
\end{figure}

We measure the effect of interactions on the tritter by comparing the outputs of an interferometer, composed of the tritter and the inverse tritter, with and without the interactions taken into account. For the purposes of our numerical calculations, we will consider the specific initial state consisting of $N$ atoms in site $0$, i.e.  $|N,0,0\rangle$.  Figure \ref{fidn} shows how the fidelity between the output states with and without the interactions varies as a function of interaction strength and for different numbers of particles, $N$.  As we would expect, the fidelity is unity when there are no interactions regardless of the value of $N$ and the effects of the interactions become more significant as $N$ is increased.  

\begin{figure}[h]
\centering
\includegraphics[scale=0.4]{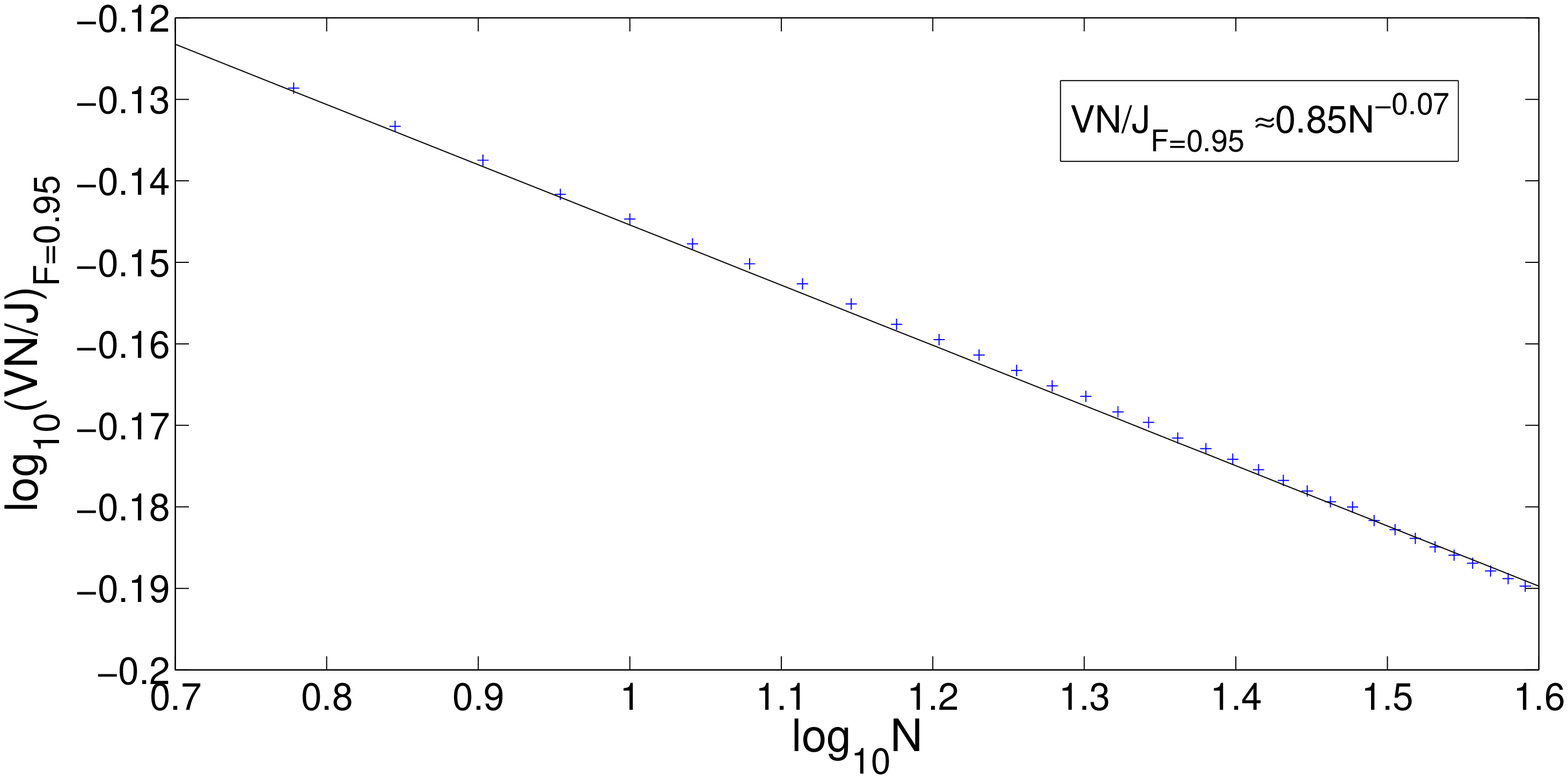}
\caption{The correlation between $VN/J$ at a critical fidelity of $F=0.95$ and the number of atoms.  The crosses are numerically calculated data points for values of $N$ up to $N=40$.  The solid curve is a line of best fit intended to find the scaling.  The scaling is, $(VN/J)_{F=0.95} \sim 0.85N^{-0.07}$ .}
\label{loginteraction}
\end{figure}

To determine the scaling of the interactions with $N$ we find the value of $VN/J$ required to give a fidelity of $0.95$ (when compared with the case without interactions) for different values of $N$.  From this we determine the relationship between the number of input atoms and $VN/J$ at this critical fidelity to be,
\begin{equation}
\left(\frac{VN}{J}\right)_{F=0.95} \approx 0.85\, N^{-0.07} 
\end{equation}
which is shown in Figure \ref{loginteraction}.  This relationship means the effect of the interactions, $V/J$, on the system scales with $N$ as $0.85N^{-1.07}$.  This is an approximately linear scaling and so the effects of interactions in the system are not too destructive.  It means we require $VN/J$ to be of the order of 0.85 for $F=0.95$.  Therefore, if we were to input $10^4$ atoms we would need $V/J \sim 10^{-4}$.
We can make $V$ very small compared to $J$ by using Feshbach resonances to tune the scattering lengths.  It is possible to achieve scattering lengths smaller than $a_0$ for some BECs \cite{Chin08} meaning we can tune $V/J$ to be of the order of $10^{-4}$, which is promising for the feasibility of the scheme.  

\begin{figure}[h]
\centering
\includegraphics[scale=0.4]{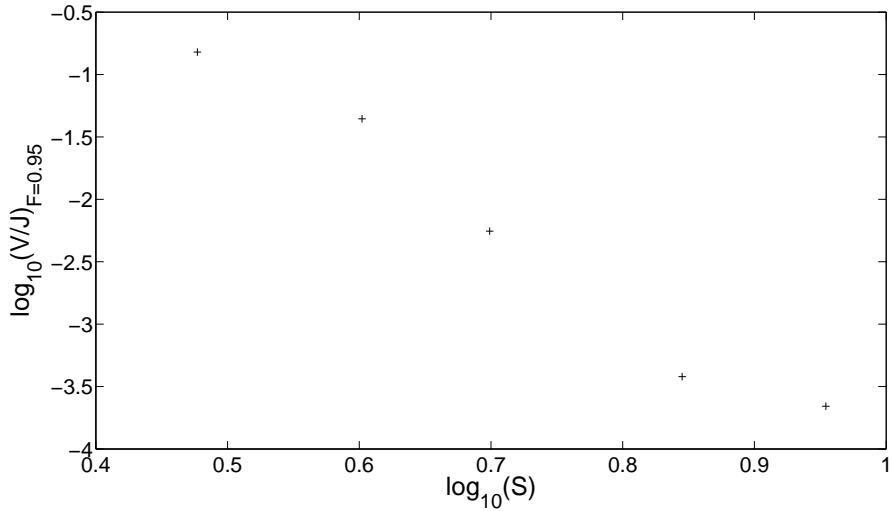}
\caption{Plot of $V/J$ at a critical fidelity of 0.95 as a function of $S$ for $N=5$.  We see $V/J$ decreases rapidly as $S$ increases. For $S=7$ the interaction strength required to achieve a fidelity of at least 0.95 is $V/J \sim 10^{-4}$.  Such a value is likely to be experimentally challenging and so we take $S=5$ as the maximum number of sites allowed when interactions are accounted for.}
\label{interactionS}
\end{figure}

Things are less promising, however, when we consider the effect of interactions in larger devices.  Determining $V/J$ at the critical fidelity of 0.95 for a fixed number of atoms ($N=5$) for $S=3, 4, 5, 7$ and $9$ allows us to see how interaction effects scale in this range of $S$.  The results are shown in Figure \ref{interactionS}.  We see that the $V/J$ that still allows us to achieve a fidelity of at least 0.95 decreases rapidly with $S$.  For $S=7$, we require $V/J \sim 10^{-4}$, which is at the limit of what is experimentally achieveable.  Interactions therefore appear to limit the current applicability of our scheme to about five sites.

\subsection{Timing errors}
\begin{figure}[h]
\centering
\includegraphics[scale=0.4]{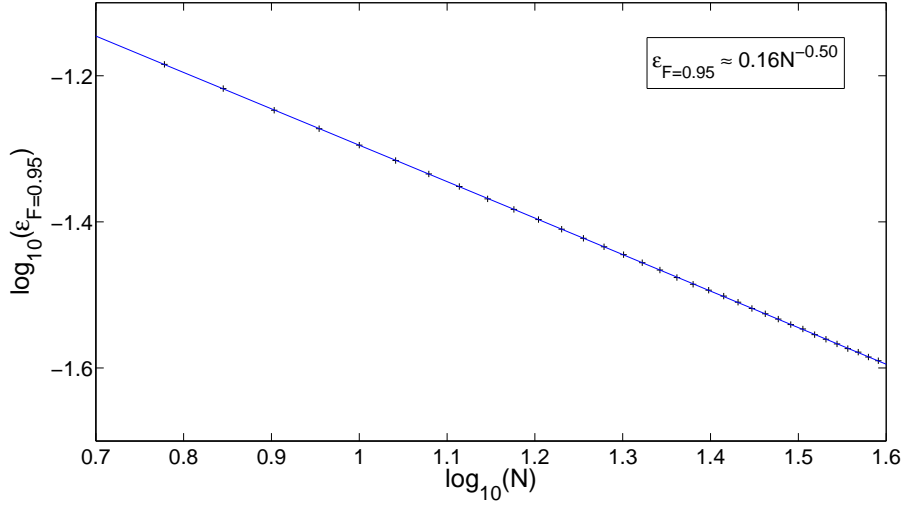}
\caption{The relationship between the absolute time error at a critical fidelity of $F=0.95$ and the number of atoms, $N$, for a tritter. The solid curve is a line of best fit and has the form $\epsilon_{F=0.95} \approx 0.16N^{-0.50}$.  We see that as the number of atoms increases, the timing error that can be tolerated decreases.}
\label{timepower}
\end{figure}

So far we have made the simplifying assumption that we can exactly measure the time the barriers are lowered for.  We will now investigate how sensitive the scheme is to inaccuracies in this timing for different numbers of atoms, $N$, and different numbers of sites, $S$.  We first consider a tritter (i.e. $S=3$). To determine the effect of an absolute time error, $\epsilon$, for different $N$ we calculate the fidelity between the outputs of the device when $Jt=\tau$ and when $Jt=\tau+\epsilon$, where $\tau$ is the $Jt$ required to produce the balanced splitter quoted in Section \ref{larger}, for the particular input state $|N,0,0\rangle$.  As with the interactions, we determine how the fidelity scales with $N$ by measuring $\epsilon$ at a critical fidelity of 0.95 for different values of $N$.  These results are shown in Figure \ref{timepower} and we see that the relationship between $\epsilon$ at the critical fidelity and the number of atoms is given by,
\begin{equation}
\epsilon_{F=0.95} \approx 0.16N^{-0.50} .
\end{equation}
The maximum time error that still allows us to achieve a fidelity of at least $0.95$ decreases relatively slowly as $N$ increases.  The same scaling of the absolute time error with $N$ was also numerically found for $S=4, 5, 7$ and $9$.

\begin{figure}[h]
\centering
\includegraphics[scale=0.4]{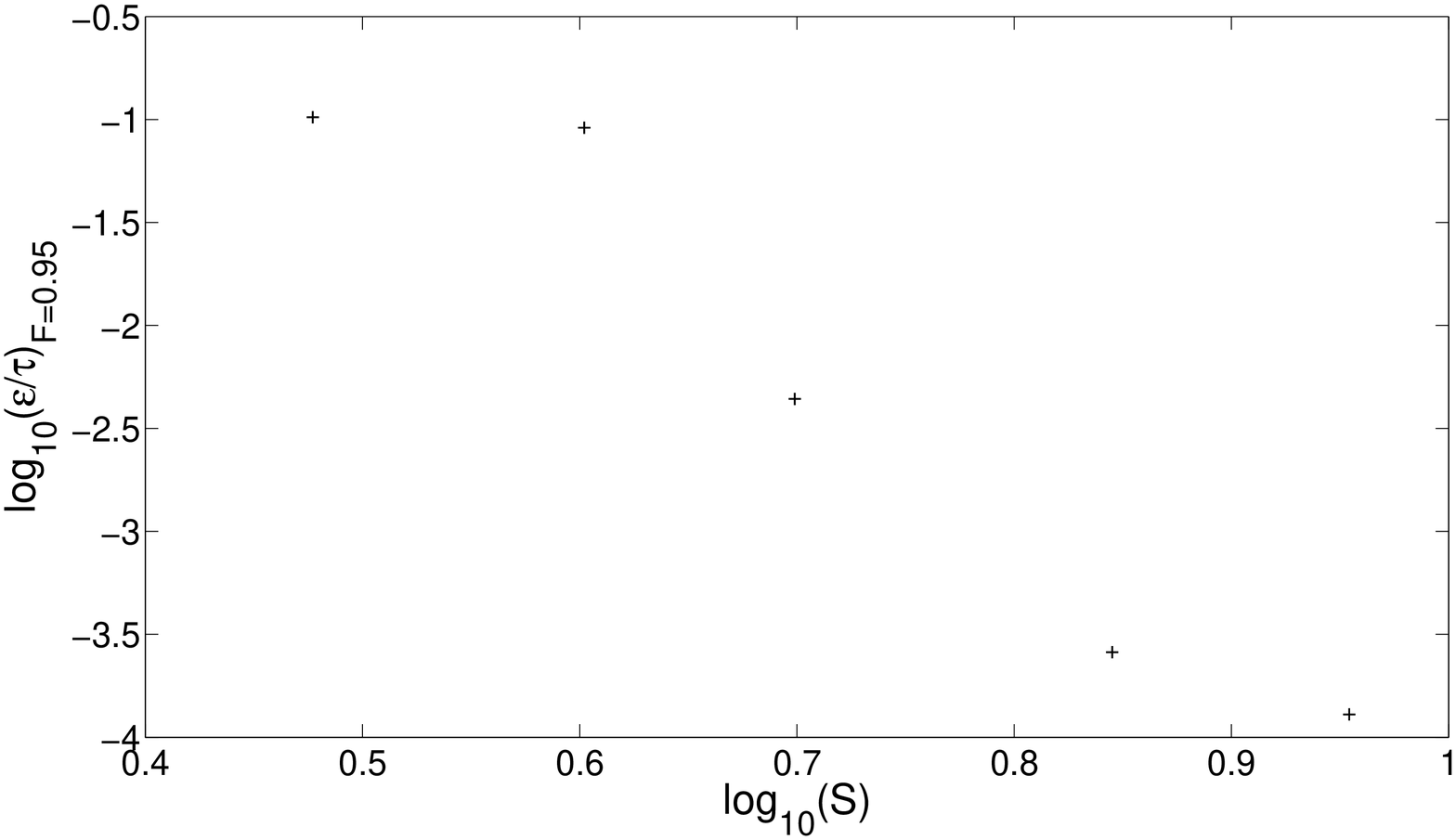}
\caption{Plot of $\epsilon/\tau$ for $S=3, 4, 5, 7$ and $9$ when we input five atoms into the system.  We determine $\epsilon/\tau$ by taking $\epsilon$ at the critical fidelity of 0.95 and dividing it by the $Jt$ required to produce balanced splitters ($\tau$) for the different $S$.  The fractional time error, $\epsilon/\tau$, that can be tolerated decreases rapidly with $S$.}
\label{scaling_timeS}
\end{figure}

However, in an experiment it is the fractional error that we must consider since clocks generate fractional errors.  We have determined these fractional errors for $S=3, 4, 5, 7$ and $9$, when we input five atoms, by dividing $\epsilon$ at $F=0.95$ by $\tau$.  These results are shown in Figure \ref{scaling_timeS}.  As expected the fractional error, $\epsilon/\tau$, that can be tolerated in these systems decreases with $S$.  For $S=9$, $\epsilon/\tau$ is of the order of $10^{-4}$.  Accuracies of one part in $10^{4}$ are easily achievable with many readily available pulse pattern generators suggesting that fractional time errors will not be the main limiting factor in this scheme.

Our scheme relies on us being in the strong coupling regime where the tunnelling rate is larger than the oscillation rate.  To ensure we satisfy this condition we take $V_0=2E_R$ and using (\ref{J}) we find $\hbar J \approx 0.06E_R$.  Substituting values for ${}^{87}Rb$ and using $\lambda \sim 1000$nm we get $J \approx 60$Hz.  We want to see how accurately we must be able to control the time given we are operating in a regime with this $J$.  From Figure \ref{timepower} we know the maximum absolute time error our system can tolerate is $\epsilon \approx 0.07$ for $N=5$ for all $S$ where $\epsilon=J \Delta t$ and $\Delta t$ is the time error in seconds.  This means we require that we can control $t$ to at least a precision given by $\Delta t = 0.07/J \approx 1$ms.  Therefore, we need a refresh frequency of about a few kilohertz which can be achieved using a ferroelectric liquid crystal spatial light modulator \cite{Fatemi06} and so we conclude that operating in the strong coupling regime does not limit our scheme.

\subsection{Rate of switching the intensity}

Another potential limit on $S$ comes about from the rate at which the barriers are raised and lowered. As discussed above, we want this to happen quickly with respect to the rate of tunnelling between the sites, $J$, but slowly with respect to the phonon excitation spectrum in each trap. This ensures that the system remains in the ground state of each trap, and any imperfection in this switching will reduce the fidelity of the output state. This separation of timescales has been experimentally demonstrated and is given by the ratio $\omega_{k}/J$, which can be calculated using Equation~(\ref{equation}).  In particular, we require $\omega_k/J >>1$ to ensure the scheme is successful.  Putting different values of $S$ into (\ref{equation}) and taking $VN/J$ to be $0.85$ we find $\omega_k/J > 1$ for all values of $S$ up to $S=12$.  This gives a limit on $S$ of 12.

\subsection{Loss of particles}

The final limiting factor we consider is spontaneous emission which we show to have little effect on the system.  In fact, when we ignore interactions, the loss of a particle during the splitting process has no effect on the remaining particles.  When $N$ particles are inputted into a tritter and one particle is lost halfway through the splitting procedure, we have numerically shown that the fidelity between this output and the output of a tritter into which $N-1$ particles are inputted (and none are lost) is 1 when the interactions are ignored ($V=0$).  Even when interactions are included the loss of fidelity is very small and so spontaneous emissions will not be the limiting factor in this scheme.

Taking all these practical limitations into account we find the limit for $S$ is five.  All that is required to produce these devices is a raising and lowering of the potential barriers and so far fewer components are used than in optical schemes to produce devices with the same number of ports.

\section{Conclusions}

We have proposed a straightforward scheme for implementing atomic beam splitters with up to five input and output ports. It requires modulation of the intensity of the optical lattice in which the atoms are trapped -- something that is readily achievable in the laboratory. Importantly, in this scheme, the multiport devices require no more operational complexity than an ordinary beam splitter for atoms.  

While experimentally challenging, the prospects for implementing these devices look promising and certainly seem to be within reach of current technologies. Indeed lattices of the required geometry have already been experimentally demonstrated \cite{Boyer2006a}. We have considered the effects of potential limiting factors and shown that, in general, they do not make the scheme prohibitively difficult.
The time the barriers are lowered for requires control to an accuracy that increases with $S$.  However, we found the precisions required to produce outputs of the required fidelity could be achieved using current laboratory equipment.  The effects of several other practical limitations were also found to be minimal.  Unfortunately, interactions were found to limit the number of ports to five.  However, the scheme is simple and has the great advantage that, for many applications, the lattice sites do not have to be addressed individually, which considerably simplifies its implementation.

This scheme is also versatile. The inverse transforms, for example, can be achieved using the same set-up just by altering the timings. In a similar way, it is possible to achieve splitters with unbalanced outputs or variable `reflectivities'.
It should also be straightforward to combine multiport devices into useful schemes such as multipath interferometers, and the `ring' lattice means they have great potential for use in gyroscopes that can measure angular momentum with Heisenberg-limited precision.
The versatility of these devices means that they are likely to have intriguing prospects for creating interesting entangled states and play a useful role in a range of new quantum technologies.

\section{Acknowledgements}
This work was supported in part by a United Kingdom EPSRC through an Advanced Research Fellowship GR/S99297/01 and the EuroQUASAR programme EP/G028427/1, a RCUK Fellowship, and a University of Leeds Doctoral Scholarship.

\newpage

\begin{thebibliography}{20}

\bibitem{Reck1994a} M. Reck, A. Zeilinger, H. J. Bernstein, and P. Bertani, Phys. Rev. Lett. {\bf 73}, 58 (1994).

\bibitem{Vourdas2004a} A. Vourdas and J.A. Dunningham, Phys. Rev. A {\bf 71}, 013809 (2005).

\bibitem{Zhang2006a} S. Zhang, C. Lei, A. Vourdas, and J. A. Dunningham, J. Phys. B {\bf 39}, 1625 (2006).

\bibitem{Mattle1995a} K. Mattle, M. Michler, H. Weinfurter, A. Zeilinger, M. Zukowski, Appl. Phys. B {\bf 60} S111 (1995).

\bibitem{Pryde2003a} G.J. Pryde and A.G. White, Phys. Rev. A {\bf 68} 052315 (2003).

\bibitem{Rasmussen94}  T. Rasmussen, A. Bjarklev and J. H. Povlsen, Electronics Lett. \textbf{30}, 583 (1994).

\bibitem{Denschlag03} J. Denschlag, \textit{et al.}, Science \textbf{287}, 97 (2000).

\bibitem{Oosten2001a} D. van Oosten, P. van der Straten, and H.T.C. Stoof, Phys. Rev. A {\bf 63}, 053601 
(2001). 

\bibitem{Rey2003a} A.M. Rey, K. Burnett, R. Roth, M. Edwards, C.J. Williams and C.W. Clark, J. 
Phys. B {\bf 36}, 825 (2003). 

\bibitem{greiner2} M. Greiner {\it et al.}, Nature (London) {\bf 419}, 51 (2002).

\bibitem{feshbach} S.L. Cornish, N.R. Claussen, J.L. Roberts, E.A. Cornell, C.E. Wieman, Phys. Rev. Lett. {\bf 85}, 1795 (2000).

\bibitem{Boyer2006a} V. Boyer, R.M. Godun, G. Smirne, D. Cassettari, C.M. Chandrashekar, A.B. Deb, Z.J. Laczik, and C.J. Foot, Phys. Rev. A {\bf 73}, 031402(R) (2006).

\bibitem{Amico2005a} L. Amico, A. Osterloh, and F. Cataliotti, Phys. Rev. Lett. {\bf 95}, 063201 (2005). 


\bibitem{Dunningham2004a} J. A. Dunningham and K. Burnett, Phys. Rev. A {\bf 70}, 033601 (2004).

%
%
%
%
%

\bibitem{Esteve08} J. Esteve, C. Gross, A. Weller, S. Giovanazzi and M. K. Oberthaler, Nature \textbf{455}, 1216 (2008).

\bibitem{Albiez05} M. Albiez, R. Gati, J. F\"olling, S. Hunsmann, M. Cristiani and M. K. Oberthaler, Phys. Rev. Lett. \textbf{95}, 010402 (2005).

\bibitem{Scheel06} S. Scheel, J. Pachos, E. A. Hinds and P. L. Knight, Lect. Notes Phys.\textbf{689}, 47 (2006).

\bibitem{Jona03} M. Jona-Lasinio, O. Morsch, M. Cristiani, N. Malossi, J. H. M\"uller, E. Courtade, M. Anderlini, and E. Arimondo, Phys. Rev. Lett. \textbf{91}, 23 (2003).

\bibitem{Kwon03} T. Y. Kwon, T. Kurosu, Y. Koga, S. Ohshima and T. Ikegami, Jpn. J. Appl. Phys. \textbf{42}, 924 (2003).

\bibitem{Chin08} C. Chin, R. Grimm, P. Julienne and E. Tiesinga, arXiv:0812.1496v1 (2008).

\bibitem{Fatemi06} F. K. Fatemi and M. Bashkansky, Opt. Lett. \textbf{31}, 7 (2006).




\end{thebibliography}
\end{document}